\documentclass[prd,aps,amsmath,twocolumn,superscriptaddress,amssymb,reprint,nofootinbib,preprintnumbers]{revtex4-2}
\pdfoutput=1
\usepackage{graphicx,array}
\usepackage{hyperref}
\usepackage{color}
\usepackage[T1]{fontenc}
\usepackage{amsmath,amssymb,slashed,latexsym}
\usepackage{bm}
\usepackage{braket}
\usepackage{placeins}
\usepackage{adjustbox}
\usepackage{enumitem}
\usepackage{slashed}

\def\beq{\begin{equation}}
\def\eeq{\end{equation}}
\def\bea{\begin{eqnarray}}
\def\eea{\end{eqnarray}}

\usepackage{array}
\newcolumntype{P}[1]{>{\centering\arraybackslash}p{#1}}
\newcolumntype{M}[1]{>{\centering\arraybackslash}m{#1}}
\usepackage{tabularray}
\UseTblrLibrary{booktabs}

\definecolor{lightblue}{rgb}{0.1, 0.5, 1.0}
\definecolor{darkblue}{cmyk}{1,0.4,0,0.3}
\definecolor{violet}{cmyk}{0,1,0,0.2}
\hypersetup{colorlinks, bookmarksnumbered, citecolor=darkblue, linkcolor=darkblue, pdfstartview=FitH, urlcolor=darkblue, linktocpage}

\begin{document}

\preprint{MIT-CTP/5750}

\title{Spontaneous symmetry breaking, gauge hierarchy and electroweak vacuum metastability}

\author{\large Sean Benevedes}\email{seanmb@mit.edu}
\affiliation{%
Department of Physics, Massachusetts Institute of Technology, Cambridge, MA 02139, USA }

\author{\large Thomas Steingasser}
 \email{tstngssr@mit.edu}
\affiliation{%
Department of Physics, Massachusetts Institute of Technology, Cambridge, MA 02139, USA }
\affiliation{Black Hole Initiative at Harvard University, 20 Garden Street, Cambridge, MA 02138, USA
}%

\author{\large Sokratis Trifinopoulos}\email{trifinos@mit.edu}
\affiliation{Department of Physics, Massachusetts Institute of Technology, Cambridge, MA 02139, USA }

\begin{abstract}
The so-called metastability bound asserts that an unnaturally small Higgs mass is a necessary condition for electroweak vacuum metastability, offering a new approach towards solving the hierarchy problem. So far, this result relies on the assumption of a negative Higgs mass parameter, or equivalently, on electroweak spontaneous symmetry breaking. We derive a new, corresponding bound for the case of a positive mass parameter. When the new bound is significantly more restrictive than or comparable to its established counterpart, it may offer an explanation for the sign of the Higgs mass parameter, and thus, spontaneous symmetry breaking itself. New physics at scales $\mathcal{O} (1-10)$~TeV can lower these bounds as far as the TeV-scale. As an illustration, we consider vacuum stability in the presence of additional TeV-scale fermions with Yukawa couplings to the Higgs, as well as a dimension-six term parameterizing new physics in the UV. This scenario requires new physics that couples strongly to the Higgs, and can potentially be probed at future colliders. Finally, to allow for comparison with concrete mechanisms predicting metastability, we provide the mass-dependent lifetime of the electroweak vacuum for this model.

\end{abstract}

\maketitle

\section{Introduction}
\label{sec:intro}

From the electron self-energy to the prediction of the $\rho$ meson, naturalness has been the paradigm of particle physics since its advent~\cite{Nelson:1985, Giudice:2008bi, Giudice:2013yca, Wells:2013tta, Wells:2016luz, Wells:2018sus, Wells:2018yyb, Wells:2021zdp, Craig:2022eqo}. This idea has, however, been challenged by the measurements of the Large Hadron Collider (LHC), and the measured value of the Higgs (pole) mass, $M_H \sim 125$~GeV. As a scalar, the natural value of the Higgs' mass should lie relatively close to the scale of new physics $\Lambda$. On the other hand, no signs of such new physics have been found at the LHC, which has been operating at energies up to $14$~TeV. At face value, this would suggest a significant fine-tuning of the parameters of any UV completion of the SM, leading to the \textit{hierarchy problem}.\footnote{The idea of such a hierarchy is consistent with a renormalization group (RG) analysis of the Standard Model (SM), which suggests perturbativity and unitarity essentially up to the Planck scale, thus allowing for an arbitrarily large $\Lambda$.}

Another apparent fine-tuning can be observed for the quartic self-coupling $\lambda$. Measurements taken at the LHC indicate that the SM electroweak vacuum exists in a delicate island of metastability~\cite{Elias-Miro:2011sqh,Degrassi:2012ry,Buttazzo:2013uya, Andreassen:2017rzq, Steingasser:2022yqx} - i.e., the Higgs potential develops a second, lower-lying minimum at high energies into which the Higgs field can tunnel, but only with a sufficiently low rate to ensure its existence over cosmological time scales. This behavior is highly sensitive to the value of essentially all SM couplings, in particular $\lambda$ itself, but also the strong gauge and top Yukawa coupling.\footnote{This sensitivity can be seen whether couplings are evaluated in the IR or in the UV; see Refs.~\cite{Steingasser:2024hqi,Buttazzo:2013uya} for further discussion.} 

Understanding the parameters in the Higgs potential therefore requires explaining these two apparent fine-tunings. In Ref.~\cite{Khoury:2021zao} it was suggested to relate these two properties - if one implies the other, a single mechanism might offer an explanation for both of them. Indeed, the authors have shown that metastability of the electroweak vacuum, in combination with some further conditions and results first presented in Ref.~\cite{Buttazzo:2013uya}, implies an upper bound on the Higgs mass orders of magnitude smaller than its natural value. This \textit{metastability bound} relies crucially on the assumption of Higgs-induced electroweak spontaneous symmetry breaking (SSB) - that is, the sign of the Higgs mass parameter $m^2$. Thus, while this bound provides a hierarchy for a negative mass term, it would still allow for a natural positive value of $m^2$, i.e., $m^2\sim\mathcal{O}(\Lambda^2)$.

In this article, we present a novel metastability bound for the case of a positive $m^2$. The combination of this upper bound with the existing lower bound from Ref.~\cite{Khoury:2021zao} ensures a hierarchy between $m^2$ and $\Lambda^2$, without relying on the assumption of SSB. Furthermore, we show that this bound can be improved by additionally demanding a finite lifetime of the vacuum, with shorter lifetimes corresponding to smaller values of the Higgs mass. We demonstrate this numerically for a concrete example of a simple Beyond Standard Model (BSM) scenario.

The novel upper bound on $m^2$ significantly expands the established connection between metastability and hierarchy. Additionally, it suggests the possibility that SSB itself may be a consequence of metastability: If the range of allowed positive values of $m^2$ is significantly smaller, or even just comparable, to the range of allowed negative values, SSB could be understood as a \textit{generic} property of metastable vacua - or at least not a surprising one. 

Our results, together with those of Ref.~\cite{Khoury:2021zao}, then suggest the possibility that the Higgs potential could be understood entirely by explaining the metastability of the electroweak vacuum. This possibility is motivated by the idea of \textit{dynamical vacuum selection}~\cite{Dvali:2003br, Dvali:2004tma, Graham:2015cka, Kaplan:2015fuy, Espinosa:2015eda, Hook:2016mqo, Arkani-Hamed:2016rle, Huang:2017oab, Geller:2018xvz, Cheung:2018xnu, Giudice:2019iwl, Arkani-Hamed:2020yna, Csaki:2020zqz, Strumia:2020bdy, Khoury:2021zao, TitoDAgnolo:2021pjo, Giudice:2021viw, Trifinopoulos:2022tfx,Giudice:2021viw, Khoury:2019ajl,Khoury:2021grg}. In these approaches, the parameters of the Higgs potential are assumed to be neither fundamental parameters nor directly related to such, but rather the result of some dynamical process. For a concrete example of how this picture could explain the metastability of the electroweak vacuum, consider Refs.~\cite{Kartvelishvili:2020thd, Khoury:2019ajl,Khoury:2019yoo,Khoury:2021grg}.

For the metastability bounds to be able to explain the observed Higgs mass, the \textit{instability scale} - i.e., the scale at which the Higgs quartic coupling $\lambda$ becomes negative due its RG running - needs to be lowered to a few TeV. Generically, this requires the presence of TeV-scale physics beyond the SM. This mechanism is thus extremely predictive; its ability to explain the observed value of the Higgs mass requires new physics near the TeV-scale, and this new physics must have a substantial coupling to the Higgs, making it testable at a future collider. As a particular example, we will consider right-handed neutrinos (RHNs) realizing a low-scale seesaw mechanism. 

The rest of this article is structured as follows. In Sec.~\ref{sec:derive_analytic_bounds}, we derive the analytic metastability bound on $m^2$ from the one-loop effective potential for the Higgs. Then, in Sec.~\ref{sec:analytic_bounds}, we describe how the metastability bounds can be lowered to bring them closer to the observed electroweak scale through the effects of BSM physics, using a simple realization of right-handed neutrinos for concreteness\footnote{We consider the same model as Ref.~\cite{Khoury:2021zao} to allow for comparisons with the lifetime computations therein.}, and present the analytic results for the bounds on $m^2$ in this model. In Sec.~\ref{sec:lifetime_constraints}, we present further results from numerical computations of the vacuum lifetime in this model, before concluding in Sec.~\ref{sec:conclusions}. Appendix~\ref{app:lifetime_calc} contains the details of our numerical lifetime computations.

\section{Analytic Metastability Bounds}
\label{sec:derive_analytic_bounds}
We derive a general, analytic upper bound on the Higgs mass in a metastable vacuum for the case $m^2 > 0$. Just as for the analysis in Ref.~\cite{Khoury:2021zao}, we represent the effects of the UV physics relating to the hierarchy problem through a dimension-six term $\propto H^6$. In addition to assuming metastability, we also assume this term to be sufficient for our analysis, i.e., that no higher-order terms affect our results. This means, in particular, that we do not consider scenarios where the Higgs potential appears to be stable in the IR, but an interplay of irrelevant operators creates a deeper minimum outside the regime of validity of the EFT. As a corollary, this implies that the Wilson coefficient of the dimension-six term, $C_6$, must be positive so that the Higgs potential is bounded from below. 

One further subtlety is that, unlike in the analysis of Ref.~\cite{Khoury:2021zao}, this assumption does not by itself imply that a second vacuum results from the quartic $\lambda$ changing sign. One can see that with a positive value of $m^2$, two vacua can result even if $\lambda$ is negative in both the IR and UV. For simplicity, we will assume that this is not the case, and that $\lambda > 0$ in the IR. This assumption is realized in the RG trajectory of $\lambda$ in the SM, and in its extensions that we consider. To satisfy this assumption and to reduce the number of free parameters, we will henceforth fix the values of all SM couplings to their observed values at the top mass scale.

A necessary condition for metastability is the existence of two minima in the (RG-improved, effective) potential. In unitary gauge, where the Higgs doublet $\mathbf{H}$ can be parameterized as $\mathbf{H} = \left(0, H/\sqrt{2} \right)$, we take the effective potential to be of the form
\begin{equation}
\label{eq:effpot}
    V_\text{eff} = \frac{1}{4}m_{\rm{eff}}^2 (H,\mu) H^2 + \frac{1}{4}\lambda_{\rm eff}(H,\mu) H^4 + \frac{C_6}{\Lambda^2}H^6\,.
\end{equation}
For $m_{\rm{eff}}^2<0$, the absolute value of this parameter coincides with the running Higgs boson mass after SSB, whereas for $m_{\rm{eff}}^2>0$, an additional factor of $2$ needs to be taken into account. Both $m_{\rm eff}^2$ and $\lambda_{\rm eff}$ are parameters of the low-energy effective field theory, i.e., include possible threshold corrections at the matching scale $\Lambda$.

\begin{figure}
    \includegraphics[width=\columnwidth]{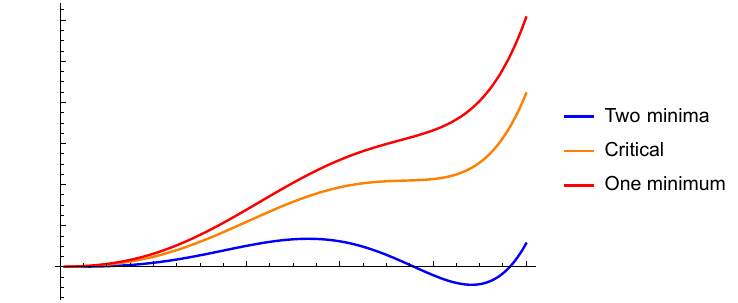}
    \caption{\label{fig:positive_mass_bound_toy} A sketch of the Higgs potential for three different (positive) values of $m^2$. For small values of $m^2$ (blue line), the potential admits a second minimum away from $H=0$, and for small enough values of $m^2$, this can be a deeper, true vacuum minimum. Large values of $m^2$ (red line) only have one, electroweak-symmetry-preserving, minimum at $H=0$. There is a critical value of $m^2$ separating these regimes, where the maximum of the potential barrier and the electroweak-symmetry-breaking minimum meet to become a saddle point.}
\end{figure}

When evaluated near the electroweak scale and for $m^2 > 0$, this potential has a unique minimum at $H=0$. As both $m^2$ and $C_6$ run multiplicatively, and $\lambda$ is taken to be positive in the IR, metastability requires that $\lambda$ changes its sign. This implies also the existence of an \textit{instability scale} $\mu_I$, where $\lambda_{\rm eff}(\mu_I, \mu_I)=0$. This allows us to rewrite this RG-improved parameter as  
\begin{equation}
    \lambda_{\rm eff}(H, \mu_I) \simeq  \beta_{\lambda} \log \left( \frac{H}{\mu_I} \right)\,.
\end{equation}
While $\lambda$ turning negative is a necessary condition for the existence of a second minimum, it is not sufficient. This can be understood through the asymptotic behavior of the potential: For values of $H$ much larger than any scale in the potential, the latter is dominated by the dimension-six term, and in the opposite limit, it is dominated by the quadratic term. For a large enough mass term, these two regimes intersect, corresponding to a strictly monotonically rising potential. Thus, for a second minimum to form, the mass parameter needs to be sufficiently small to allow for a regime in which the quartic term becomes dominant. For this to give rise to a second minimum, $\lambda$ needs to be negative in this regime (see Fig.~\ref{fig:positive_mass_bound_toy}).

On a quantitative level, these two cases can be distinguished by the number of local extrema of the potential, i.e., solutions to the equation
\begin{equation}\label{eq:extrema}
    0 = m^2 - 2 |\beta_\lambda| \left( \log\left(\frac{H}{\mu_I}\right) + \frac{1}{4} \right)H^2 + \frac{12 C_6}{\Lambda^2} H^4\,.
\end{equation}
For a large mass parameter, there are no non-zero solutions. For a small mass parameter, there can be two such solutions, as required by metastability. For the \textit{critical value} $m_+^2$ of the mass separating these regimes, there exists one solution, representing a saddle point. Here, as in the following, the index ``+'' indicates the sign of the mass term.

To determine the number of solutions, we use the technique described in great detail in Ref.~\cite{Steingasser:2023ugv}. We first simplify Eq.~\eqref{eq:extrema} by introducing the auxiliary variable $x$ as
\begin{equation}
    x \equiv \frac{H^2}{\mu_I^2}\,.
\end{equation}
This leads to
\begin{equation}
    \frac{m^2}{\mu_I^2 |\beta_\lambda|} =  x \left( \log x + \frac{1}{2} \right) - \hat{C}_\Lambda x^2 \equiv F(x)\,, 
\end{equation}
where
\begin{equation}
    \hat{C}_\Lambda \equiv \frac{12 C_6 \mu_I^2}{\Lambda^2 |\beta_\lambda|}\,.
\end{equation}
The function $F(x)$ is bounded from above, such that this equation has precisely one solution iff its left hand side is equal to the maximum possible value of $F(x)$. $F(x)$ is maximized for
\begin{equation}
    x_{\rm{max}} = -\frac{1}{2 \hat{C}_\Lambda} W_{-1} \left( -2 \hat{C}_\Lambda e^{-3/2} \right)\,,
\end{equation}
where $W_{-1}$ denotes the $k=-1$ branch of the Lambert $W_k$ function, and $e$ is Euler's number. 

This results in a bound on the Higgs mass of
\begin{equation}
\label{eq:analytic_mass_bound}
    m^2 \leq \frac{|\beta_\lambda|^2 \Lambda^2}{48 C_6} W_{-1} \left(\xi \right) \left(2 + W_{-1} \left(\xi \right)\right) \equiv m_+^2\,,
\end{equation}
where $\xi \equiv -2 \hat{C}_\Lambda e^{-3/2}$. 

The right-hand side of the inequality is not bounded from below, whereas, by assumption, $m^2 > 0$. Demanding that $m_+^2 > 0$ then leads to an additional bound,
\begin{equation}
\label{eq:analytic_mui_bound}
   \mu_I^2  <  \frac{|\beta_\lambda|}{12 \sqrt{e} \cdot C_6} \cdot \Lambda^2\,.
\end{equation}
This agrees with our earlier arguments: If the instability scale is too large relative to $\Lambda$, $\lambda$ only becomes negative once the positive dimension-six term is already dominant, and no second minimum is generated.

For generic values of $\xi$, it is reasonable to consider $W_{-1}(\xi)$ as $\mathcal{O}(1)$, so that the bound is parametrically
\begin{equation}
    m^2 \lesssim |\beta_\lambda|^2 \Lambda^2\,.
\end{equation}
Therefore, metastability generally requires $m^2$ to be suppressed by \textit{at least} one loop order relative to $\Lambda$. For values of $\xi$ close to saturating Eq.~\eqref{eq:analytic_mui_bound}, the bound is further enhanced relative to this parametric estimate.

In Ref.~\cite{Khoury:2021zao}, a similar bound has been derived for the case $m^2<0$,
\begin{equation}\label{eq:lowerbound}
     |m^2| <  m_-^2 \sim  |\beta_\lambda| \mu_I^2 \ll |\beta_\lambda| \Lambda^2\,. 
\end{equation}
The combination of Eq.~\eqref{eq:analytic_mass_bound} and Eq.~\eqref{eq:lowerbound} then bounds $m^2$ from above and from below, ensuring a hierarchy between $m^2$ and $\Lambda$ independent of the sign of the mass term. 

\section{Results for Analytic Bounds in a Low-Scale Seesaw Model}
\label{sec:analytic_bounds}

The numerical values of $m_+^2$ and $m_-^2$ are ultimately determined by the instability scale $\mu_I$. For the central values of the SM couplings, this scale is roughly of order $\mathcal{O}(10^{11})$~GeV, compared to the observed mass parameter of $\mathcal{O}(10^2)$~GeV. This discrepancy can be remedied through the effects of new physics on the running of $\lambda$. As a well-motivated simple example, we consider right-handed neutrinos (RHNs) at the TeV-scale, following Refs.~\cite{Khoury:2021zao,Chauhan:2023pur}. If their masses are lower than the SM instability scale and they couple to the Higgs sector through sizeable Yukawa couplings, they would manifest through a negative term in the beta function of $\lambda$, driving it towards zero already at lower energies.

The observation of non-zero SM neutrino masses provides strong motivation for the presence of RHNs, which can generate naturally light masses for the SM neutrinos through the well-known seesaw mechanism~\cite{Minkowski:1977sc,Glashow:1979nm,Mohapatra:1979ia,Gell-Mann:1979vob,Yanagida:1980xy,Schechter:1980gr}. The simplest realization of this idea, usually referred to as \textit{Type-I}, is described by the Lagrangian
\begin{equation}
   \mathcal{L} _{\nu} \hspace{-0.2 em} = \hspace{-0.2 em} \Bar{N}_i i \slashed{\partial} N_i \hspace{-0.06 em} -  \hspace{-0.06 em} \bar{\mathbf{L}}_\alpha \mathbf{H}^c (Y_\nu)_{i}^{\alpha} N_i \hspace{-0.06 em} -  \hspace{-0.06 em}  \frac{1}{2}\Bar{N}_i^c (M_N)_{j}^{i} N_j + \text{h.c.}\,,
\end{equation}
where $N_i$ are the RHNs, $\mathbf{L}$ is the SM lepton doublet, $Y_\nu$ is the matrix of Yukawa couplings of the RHNs, and $M_N$ is their Majorana mass matrix. The Greek and Latin generation indices run from $1$ to $3$. For a generic structure of the mass matrix $M_N$, the left-handed neutrinos' masses arising from this theory scale as $v^2/M$, with $M$ the characteristic scale of $(M_N)_{ij}$. Naively, this seems to suggest RHNs far heavier than the SM instability scale, making it impossible for them to affect its value through their effect on the running of $\lambda$. However, this neglects the possibility of nontrivial matrix structure in $M_N$ and $Y_\nu$.

We will thus consider the above model with an approximate $B-\tilde{L}$ symmetry, as developed in Refs.~\cite{Ghiglieri:2020ulj,Canetti:2012kh,Shaposhnikov:2006nn,Shaposhnikov:2008pf,Asaka:2005an,Asaka:2005pn,Bezrukov:2012sa,Agrawal:2021dbo}. This approximate symmetry yields the small observed masses of the left-handed neutrinos with RHNs potentially at much lower scales, including $\mathcal{O}(\text{TeV})$.

The effect of the small breaking of the $B-\tilde{L}$ symmetry on the RG trajectories of couplings is negligible. When realized exactly, this symmetry fixes the structure of $(M_N)_{ij}$, which is thus fully determined by the single parameter $M$. Moreover, it was shown in Ref.~\cite{Chauhan:2023pur} that the effects of the RHNs on the running are then encompassed by the \textit{trace parameter} $T_2 \equiv \text{Tr}(Y_\nu^\dagger Y_\nu)$. Thus, the RHN sector only introduces two new parameters relevant to our analysis, $T_2$ and $M$. Our results only depend on $C_6$ through the ratio $C_6/\Lambda^2$, so that we can take $C_6 = 1$ without loss of generality and are left with a three-dimensional parameter space.

To more clearly illustrate the effects of the RHNs, we fix $\Lambda = 1000$~TeV. This value is close enough to the electroweak scale for the resultant metastability bounds to be relevant for the hierarchy problem, and high enough to allow for the required hierarchies between $m^2$, $\mu_I^2$, and $\Lambda^2$. We examine each of the metastability bounds separately to demonstrate their qualitative features before ultimately combining the bounds to constrain $m^2$ independent of its sign.

We first run the SM couplings at three-loop accuracy (with the largest four-loop contributions for $\alpha_{s}$ taken into account), using the $\beta$ functions from Ref.~\cite{Buttazzo:2013uya}, with two-loop contributions from the RHN sector from Ref.~\cite{Pirogov:1998tj}. We include SM contributions to the effective potential from the top quark and the gauge bosons, as well as the contribution from integrating out the RHN. For the initial conditions, we use the SM couplings from Ref.~\cite{Huang:2020hdv}.

\begin{figure}[t!]
    \centering
     \includegraphics[width=\columnwidth]{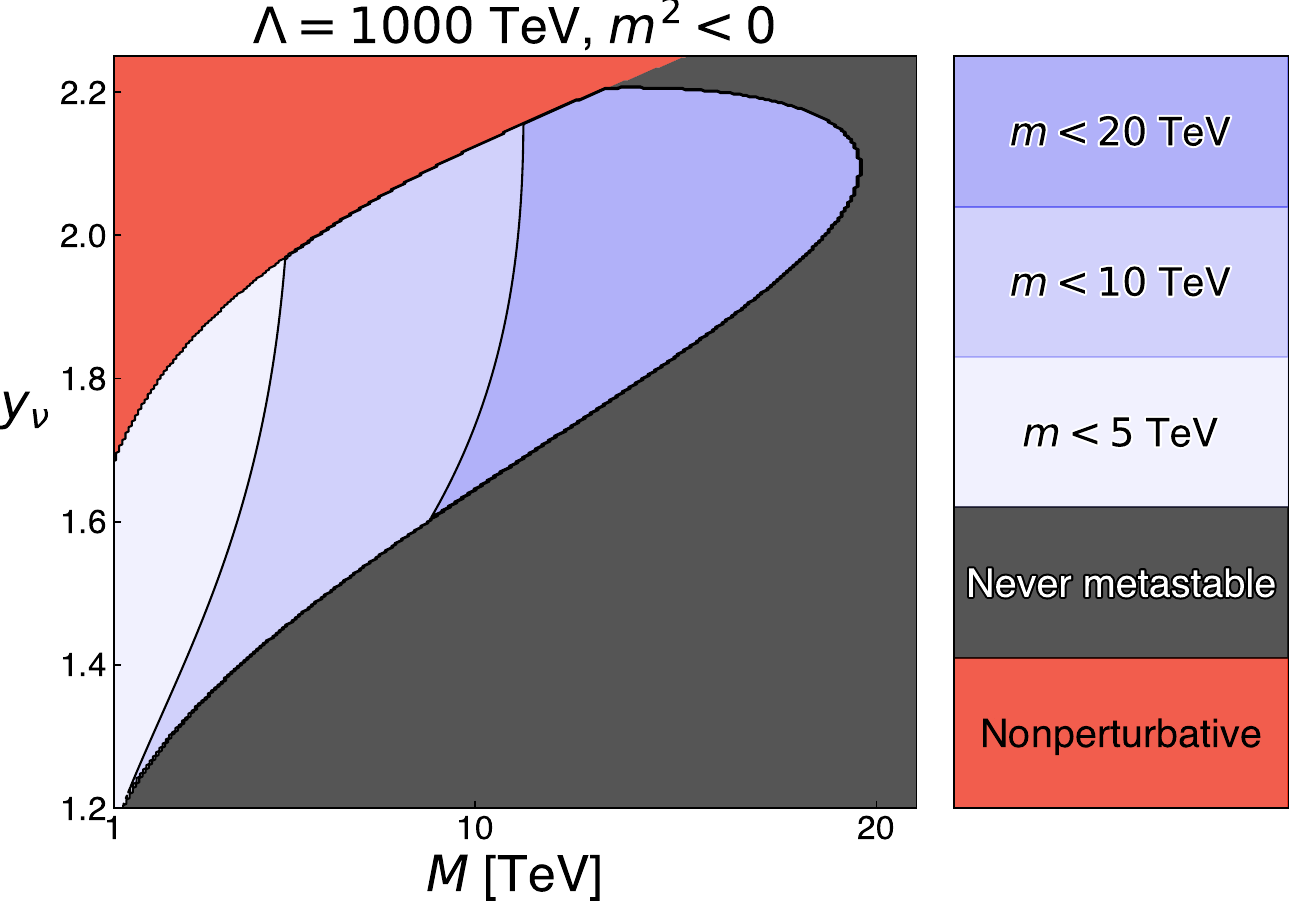}
    \caption{Maximum allowed value for $m$ on the $M-y_\nu$ plane for $m^2 < 0$, where $M$ is the RHN scale and $y_\nu \equiv \sqrt{T_2}$ parameterizes the strength of the RHN coupling to the Higgs. The red region is not considered due to couplings becoming non-perturbative before the scale $\Lambda$. The grey region is where the Higgs potential does not admit a metastable minimum even for $m^2 = 0$, so it is completely excluded by metastability. The shades of blue then show the metastability bounds on $m$, which are increasingly stringent from darker to lighter shades.}
    \label{fig:fig2}
\end{figure}

First, in Fig.~\ref{fig:fig2}, we present the bound on $m \equiv \sqrt{|m^2|}$ for $m^2 < 0$ as a function of the neutrino parameters, $M$ and $y_\nu \equiv \sqrt{T_2}$.
We find that demanding perturbativity and the existence of a metastable minimum for \textit{some} value of $m^2$ yields a finite region of viable parameter space. We further observe that, as expected, the strength of the Higgs mass bound increases as the RHN sector becomes lighter and more strongly coupled to the Higgs, because the bound depends primarily on the value of $\mu_I$.

\begin{figure}[t!]
    \centering
     \includegraphics[width=\columnwidth]{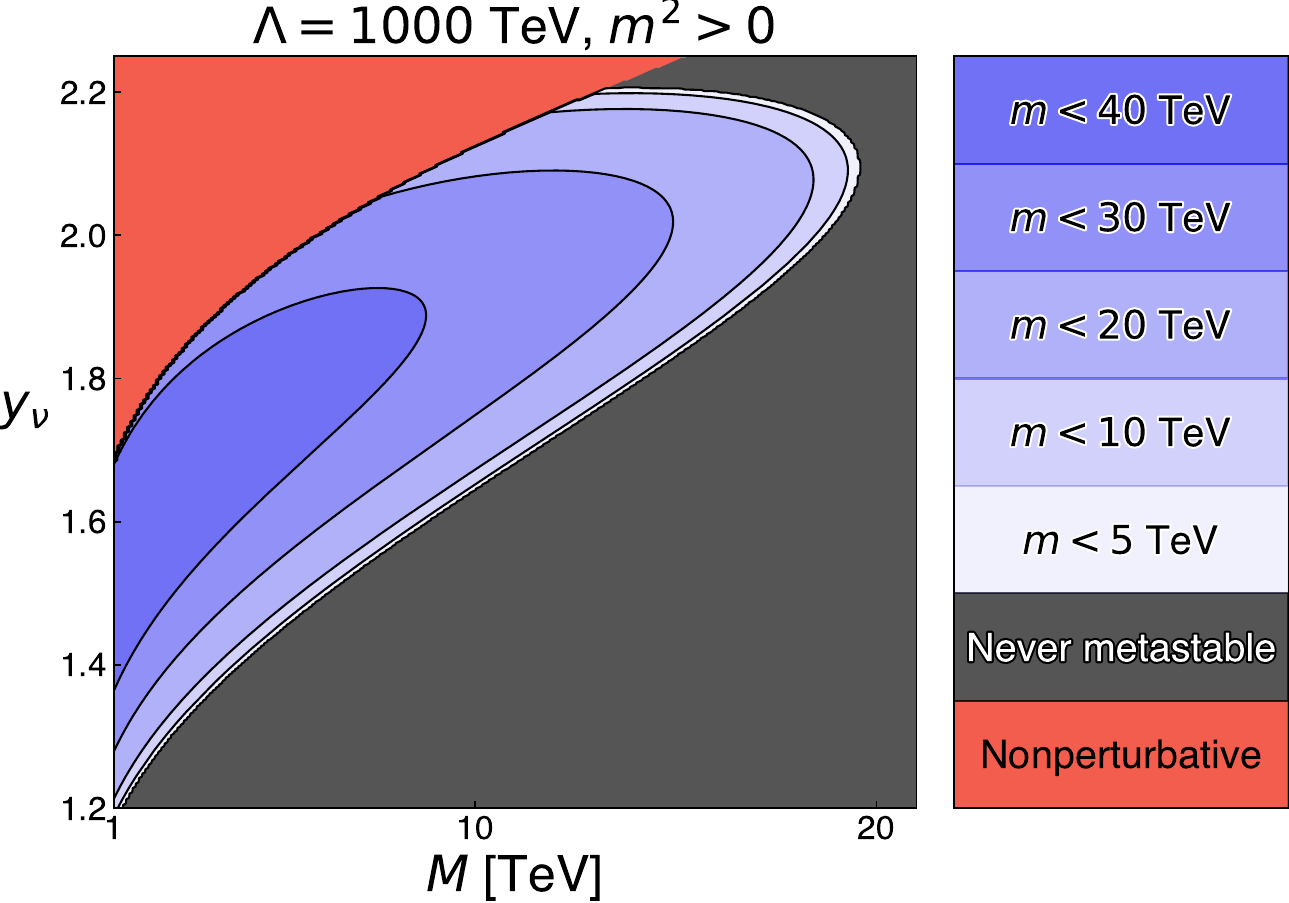}
    \caption{Maximum allowed value for $m$ on the $M-y_\nu$ plane for $m^2 > 0$, where $M$ is the RHN scale and $y_\nu \equiv \sqrt{T_2}$ parameterizes the strength of the RHN coupling to the Higgs. The red region is not considered due to couplings becoming non-perturbative before the scale $\Lambda$. The grey region is where the Higgs potential does not admit a metastable minimum even for $m^2 = 0$, so it is completely excluded by metastability. The shades of blue then show the metastability bounds on $m$, which are increasingly stringent from darker to lighter shades.}
    \label{fig:fig3}
\end{figure}

In Fig. \ref{fig:fig3}, we present the bound for $m^2 > 0$. We find that this bound is, in large parts of parameter space, comparable to, but weaker than, the $m^2 < 0$ bound. Furthermore, the bound is strongest when $M$ and $y_\nu$ are such that the two minima of the potential are nearly degenerate for $m^2 = 0$, such that even a small positive $m^2$ is sufficient to lift the high-scale minimum above the $H=0$ minimum.

\begin{figure}[t!]
    \includegraphics[width=\columnwidth]{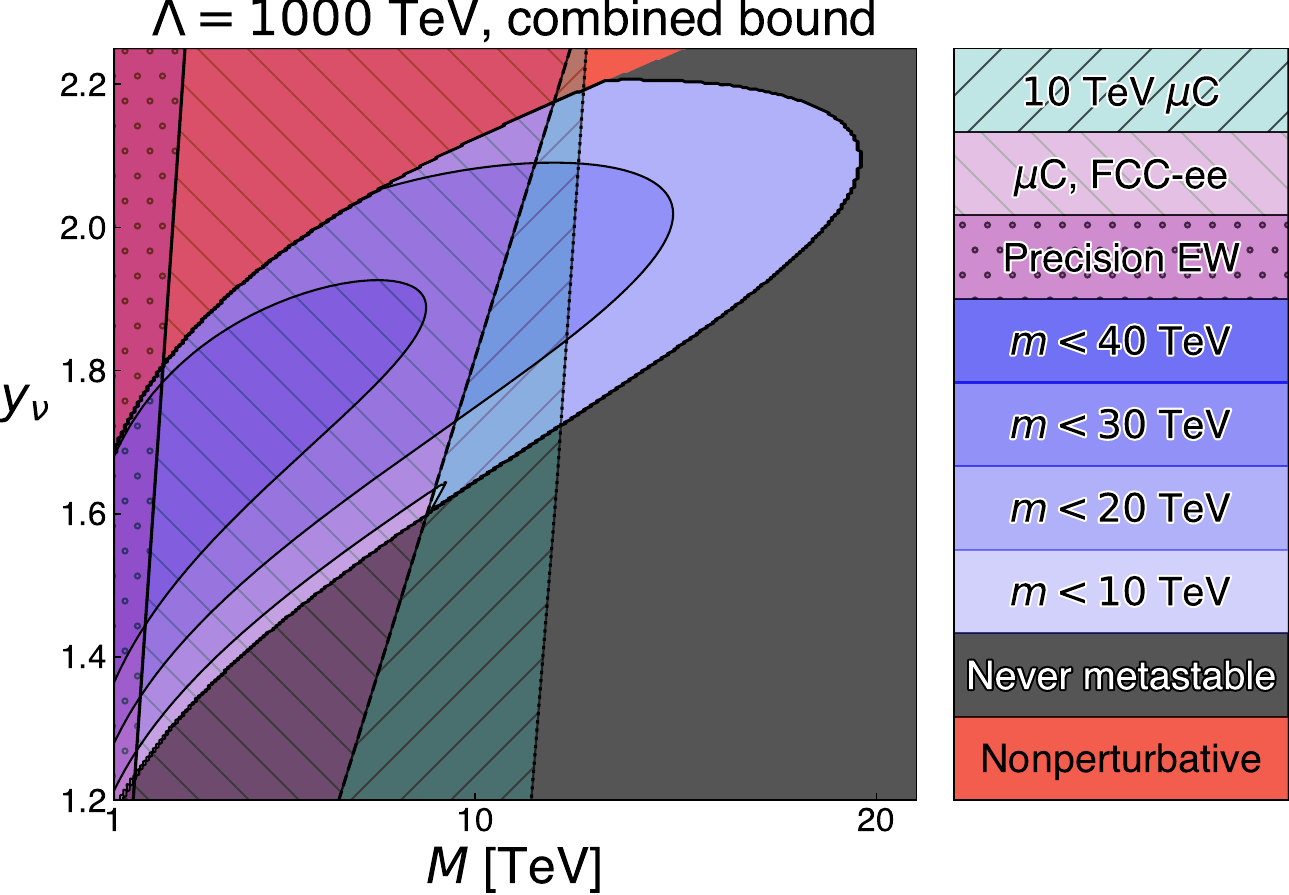}
    \caption{\label{fig:fig4}The combined bounds on $m$ in the $M - y_\nu$ plane, where $M$ is the RHN scale and $y_\nu \equiv \sqrt{T_2}$ parameterizes the strength of the RHN coupling to the Higgs. The red region is not considered due to couplings becoming non-perturbative before the scale $\Lambda$. The grey region is where the Higgs potential does not admit a metastable minimum even for $m^2 = 0$, so it is completely excluded by metastability. The shades of blue show the metastability bounds on $m$, which are increasingly stringent from darker to lighter shades. We also include relevant present constraints (dotted, adapted from Ref.~\cite{Chauhan:2023pur} originally from Refs.~\cite{delAguila:2008pw,Akhmedov:2013hec,deBlas:2013gla,Antusch:2014woa,Blennow:2016jkn,Flieger:2019eor}) and projected future searches (hashed): FCC-ee projections from Ref.~\cite{Mekala:2022cmm} and muon collider projections from Ref.~\cite{Mekala:2023diu}). The sharp corner of the $10$ TeV region is due to the discontinuous transition from the bound being $m_+^2$ dominated to $m_-^2$ dominated.}
\end{figure}

Fig.~\ref{fig:fig4} shows the combined bounds on $m$, given by the square root of $m_{\pm}^2 \equiv \text{max}(m_-^2, m_+^2)$. We overlay this bound with present experimental constraints compiled from Ref.~\cite{Chauhan:2023pur} (in turn adapted from Refs.~\cite{delAguila:2008pw,Akhmedov:2013hec,deBlas:2013gla,Antusch:2014woa,Blennow:2016jkn,Flieger:2019eor}), as well as future projected bounds from the electron-electron Future Circular Collider (FCC-ee)~\cite{Mekala:2022cmm} and from a $10$ TeV muon collider~\cite{Mekala:2023diu}. The region of parameter space in which $m$ is bounded to be less than $10$ TeV is currently not strongly constrained by experiments, but the entirety of this region can be probed by proposed future colliders (in addition to a significant fraction of the parameter space of interest as a whole).

%

\begin{figure}
     \includegraphics[width=\columnwidth]{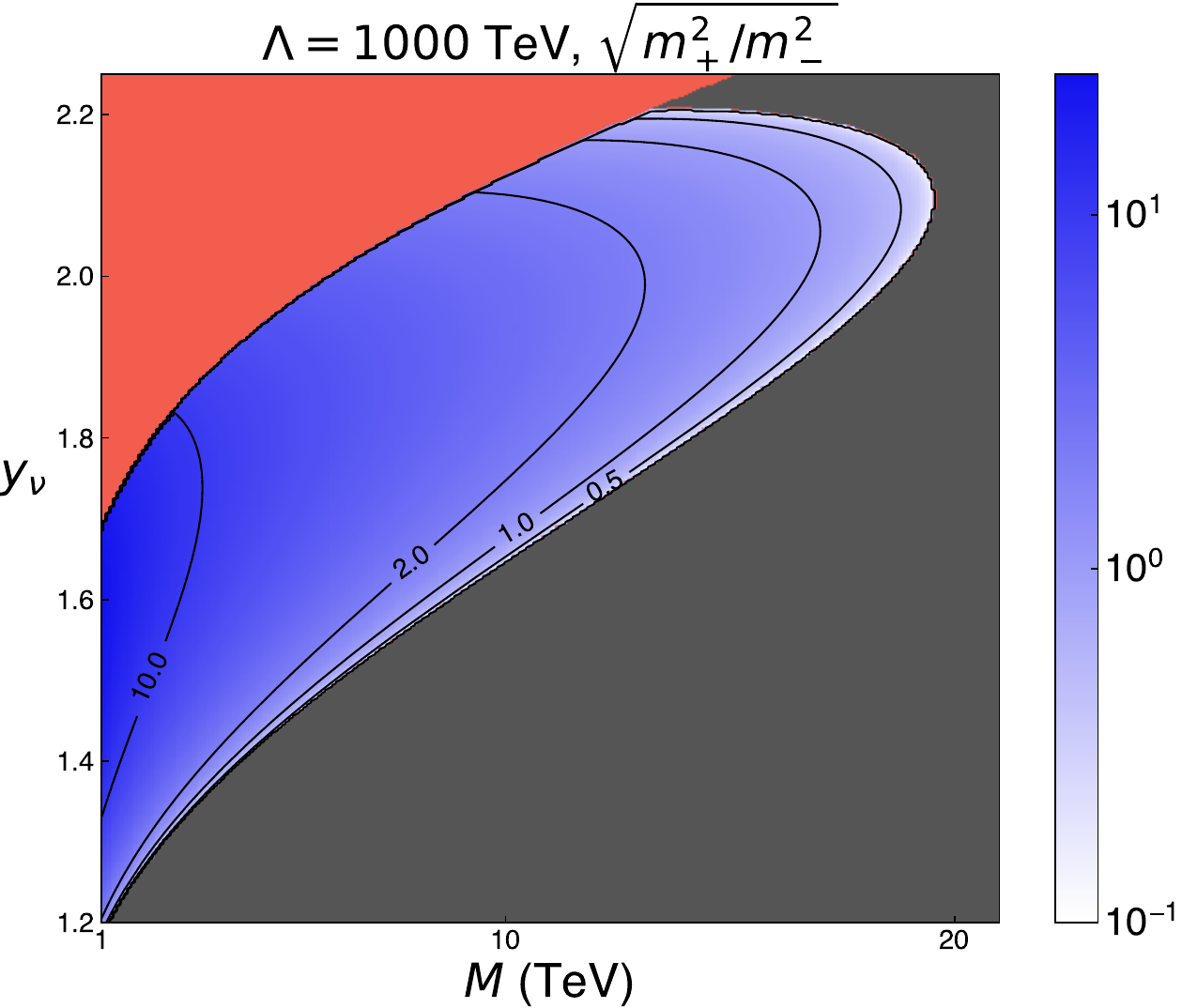}
    \caption{\label{fig:fig5} Contour plot of $\sqrt{m_+^2/m_-^2}$ in the $M - y_\nu$ plane, where $M$ is the RHN scale and $y_\nu \equiv \sqrt{T_2}$ parameterizes the strength of the RHN coupling to the Higgs. Lighter blue corresponds to stronger bounds from above and darker blue corresponds to stronger bounds from below. Contour lines are at values of $1/2, 1, 2,$ and $10$, from right to left. As before, red indicates the region where couplings run to non-perturbative values before the scale $\Lambda$, while in the grey region there is no metastable vacuum for any value of $m^2$.}
\end{figure}

Finally, we consider the relative sizes of the bounds. If the bound from below, $m_-^2$, were significantly stronger than the bound from above, $m_+^2$, the observed electroweak symmetry breaking in our universe could be understood as improbable from our IR perspective, requiring further explanation in the UV.\footnote{A definite statement about this matter would ultimately require specifying a probability measure on parameter space, which would be closely linked to the mechanism responsible for selecting the Higgs mass parameter. Henceforth, we will restrict ourselves to considering the general plausibility of certain properties, which implies the assumption of a flat prior.} And indeed, we had found before that this is the case for large parts of parameter space, see Figs.~\ref{fig:fig3} and \ref{fig:fig4}. To allow for a better comparison, we plot $(m_+^2/m_-^2)^{\frac{1}{2}}$ in Fig.~\ref{fig:fig5}. We find that the ratio is typically $\mathcal{O}(1)$, except right on the boundary between the blue and grey regions, where $m_+^2$ can be made arbitrarily strong. In this region, we obtain a strong preference for SSB. These results suggest that the presence of SSB might not be excessively improbable at any point in our parameter space.

\begin{figure}
     \includegraphics[width=\columnwidth]{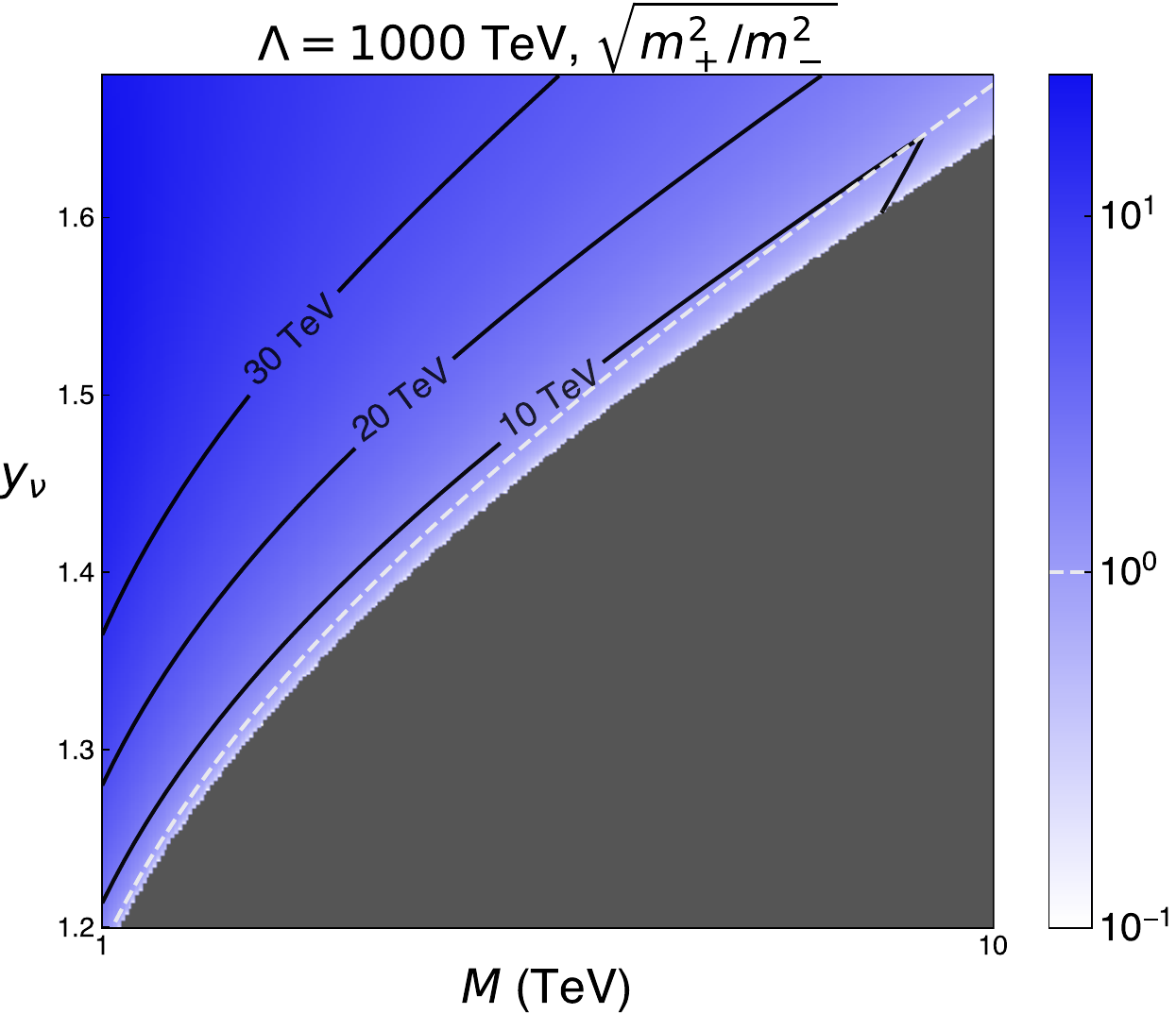}
    \caption{\label{fig:fig6} A zoom-in of Fig. \ref{fig:fig5}, $\sqrt{m_+^2/m_-^2}$ in the $M - y_\nu$ plane, where $M$ is the RHN scale and $y_\nu \equiv \sqrt{T_2}$ parameterizes the strength of the RHN coupling to the Higgs. Grey again shows regions where there is no metastable vacuum for any value of $m^2$. The solid black contour lines show the combined bounds on $m$ as shown in Fig.~\ref{fig:fig4}, and the dashed white contour line is where $m_+^2 = m_-^2$.}
\end{figure}

An important feature of the results shown in Fig.~\ref{fig:fig5} is that the region where this ratio is small is also the region where the overall bound on the Higgs mass is closest to the electroweak scale. Fig.~\ref{fig:fig6} presents an enlarged view of this region.
We find that if the Higgs mass is bound to be smaller than $10$ TeV, $m_+^2$ is smaller than $m_-^2$ for about half of the parameters, and larger but comparable in the rest. In other words, there exists a region in our RHN parameter space in which the requirement of metastability would not only be able to explain the observed value of the Higgs mass, but possibly \textit{also} its sign, and hence, SSB itself. 
\section{Lifetime constraints}
\label{sec:lifetime_constraints}
In Sec.~\ref{sec:analytic_bounds}, we have only demanded that a metastable vacuum can exist at all, while being agnostic about its lifetime. However, it is plausible that the UV dynamics responsible for metastability would prefer particular values of the vacuum lifetime $\tau$, as is indeed the case for the mechanism developed in Refs.~\cite{Kartvelishvili:2020thd,Khoury:2019yoo,Khoury:2019ajl,Khoury:2021grg}. In the following, we will show how fixing $\tau$ affects the value of $m^2$.

The vacuum decays through the formation of a ``bubble of true vacuum'', which then expands with essentially the speed of light. The lifetime can thus be defined quantitatively by demanding that the probability that such a bubble has formed within the past lightcone of any given observer is $\mathcal{O}(1)$~\cite{Kobzarev:1974cp, Coleman:1977py, Callan:1977pt}. Assuming a dark energy-dominated Universe with Hubble constant $H_0$, it was shown in Ref.~\cite{Buttazzo:2013uya} that one can relate this time to the bubble nucleation rate per unit volume through
\begin{equation}
    \tau = \frac{3 H_0^3}{4  \pi} \left(\frac{\Gamma}{V}\right)^{-1}\,.
\end{equation}
As first pointed out in Refs.~\cite{Fubini:1976jm, Lipatov:1976ny, Coleman:1977py, Callan:1977pt, Coleman:1980aw}, this rate can be brought to the form
\begin{equation}
    \frac{\Gamma}{V} \simeq A e^{-S_E[\phi_I]}\,,
\end{equation}
where $\phi_I$ is the familiar \textit{instanton}, $S_E$ is its Euclidean action, and $A$ is a dimensionful prefactor.\footnote{For a more modern perspective, see Refs.~\cite{Andreassen:2016cvx,Andreassen:2017rzq,Steingasser:2024ikl,Steingasser:2023gde}.} 

An important subtlety of this result is that the existence of an instanton is not guaranteed. An important example of a model where no such solution exists is the SM with a non-vanishing Higgs mass term~\cite{Andreassen:2017rzq}. In our case, however, it is straightforward to show that this is not an issue using the methods described in Ref.~\cite{Espinosa:2020qtq}. Further details about the instanton configuration and other aspects of the lifetime calculation are contained in Appendix~\ref{app:lifetime_calc}.

Following our reasoning in Sec.~\ref{sec:analytic_bounds}, we calculate $\tau$ taking into account the effects of RHNs with an approximate $B-\tilde{L}$ symmetry as well as a dimension-six operator. We scan over $M,~y_\nu $, and $m^2,$ while again fixing $\Lambda = 1000$ TeV. We furthermore restrict ourselves to positive values of $m^2$, referring to Ref.~\cite{Khoury:2021zao} for the case of a negative $m^2$. Since we remain agnostic to the origin of metastability, and hence, preferred lifetimes, we report our results for a variety of values for $\tau$.

In our calculation, we use the $1-$loop effective potential obtained by integrating out the heavy neutrino states, the top quark, and the vector bosons, which is justified for small $\lambda$~\cite{Gleiser:1993hf,Steingasser:2023gde}. For the running of the couplings, we repeat the procedure of Sec.~\ref{sec:analytic_bounds}. We numerically find the instanton configuration in this potential using a standard shooting procedure and compute the $1-$loop contributions from the Higgs by calculating its corresponding functional determinant using the Gelfand-Yaglom method~\cite{Gelfand:1959nq}.

\begin{figure*}
    \centering    \includegraphics[width=\textwidth]{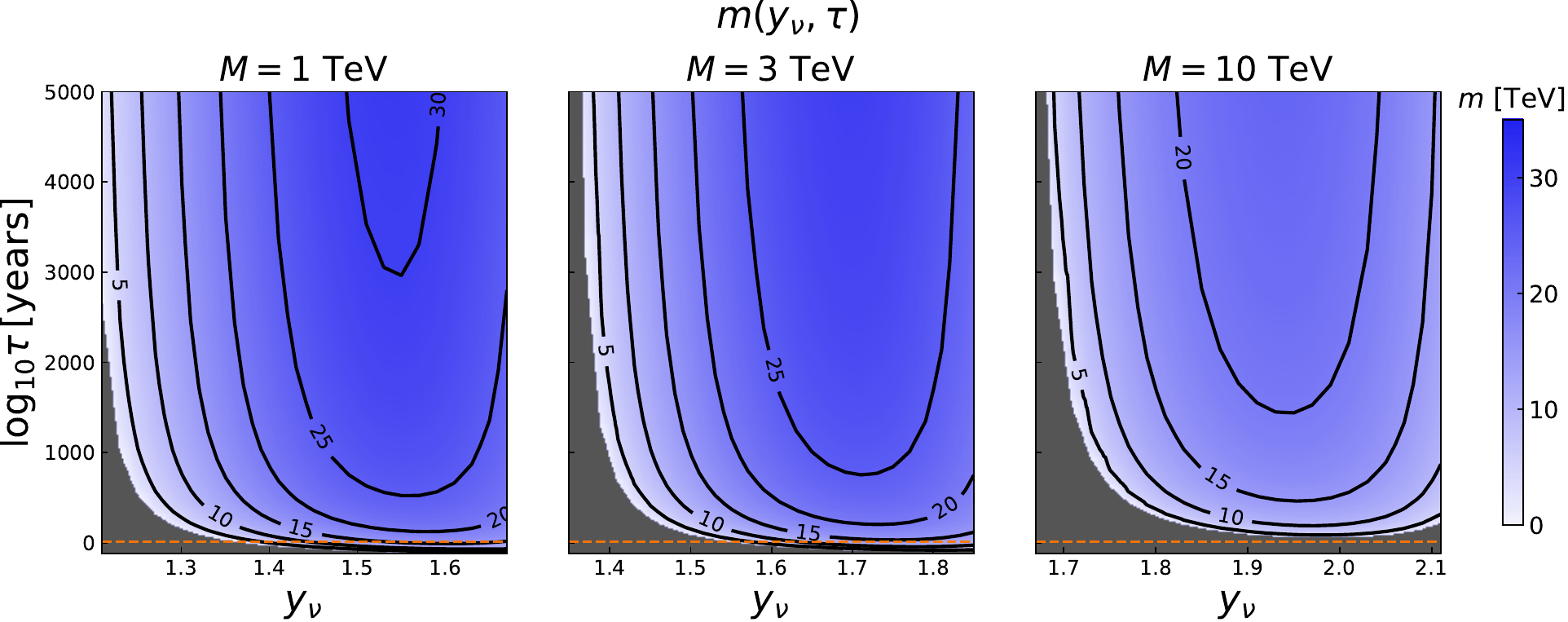}
    \caption{$m$ as a function of the logarithm of the lifetime $\tau$ in years and the RHN Yukawa $y_\nu$. The panels correspond to a RHN mass scale of $M = 1$, $3$, and $10$~TeV respectively. Shades of blue denote the value of $m$, with darker blue meaning larger $m$. Contours of $m$ are placed (with inline labels) at $5$, $10$, $15$, $20$, $25$, and $30$ TeV; not all contours are present in all panels, since the range of $m$ differs. The grey region shows values of $y_\nu$ and $\log_{10} \tau$ which do not arise for any value of $m$ (for $m^2 > 0$). The orange dashed line in each panel corresponds to the age of our universe, which we take to be $10^{10}$ years. $y_\nu$ ranges in each panel from the minimum value for metastability to the maximum value for perturbativity; $\log_{10} \tau$ ranges from the minimum value we encounter in our calculations to $5000$. We find that, for a given value of $y_\nu$, a shorter lifetime corresponds to a larger hierarchy. This qualitative behavior agrees with that of the case $m^2<0$ discussed in Ref.~\cite{Khoury:2021zao}.}
    \label{fig:fig7}
\end{figure*}

We present our results for the lifetime in Fig.~\ref{fig:fig7}, showing the value of $m$ predicted for each value of $\tau$ as a function of the RHN parameters and $\Lambda$. We find that for fixed $y_\nu$, $m$ increases monotonically with $\tau$. This is because a larger value of $m$ lifts the true vacuum relative to the false vacuum, increasing the lifetime of the false vacuum. For fixed $\tau$, the relationship between $m$ and $y_\nu$ is not monotonic. As $y_\nu$ increases, the corresponding value of $m$ required to achieve the lifetime $\tau$ increases at first, as $\lambda$ is driven negative more and more quickly. This relationship reverses when the effect of $y_\nu$ on the running of the other couplings (particularly $C_6$) becomes dominant. This behavior can also be seen on the level of the analytic bounds, see, for example, the dependence on $y_\nu$ in Fig. \ref{fig:fig3}.

\section{Conclusions}
\label{sec:conclusions}
We have shown that the requirement of electroweak vacuum metastability, in conjunction with reasonable assumptions on the SM as an effective field theory and fixed SM couplings in the IR, bounds $m^2$ with no need to assume its sign. This ensures a hierarchy between the electroweak scale and new physics at a scale $\Lambda$. Assuming validity of the SM up to the Planck scale, this bound lies far above the observed value of $m^2$, and hence cannot explain the lightness of the Higgs. 

However, in models with new physics at the TeV-scale, such as the low-scale seesaw mechanism that we have considered, the metastability bound on $m^2$ can be tight enough to be relevant for explaining the observed value of the electroweak scale. This requires new physics not far above the TeV-scale with intermediate-to-strong couplings, which can be probed at future experiments even if the UV physics responsible for metastability remains far out of their reach. Moreover, detection of such TeV-scale particles destabilizing the Higgs potential would immediately imply the existence of new physics at a higher UV scale $\Lambda$ to prevent too rapid vacuum decay. 

In this framework, the task of solving the hierarchy problem through these means is equivalent to justifying the assumptions leading to the metastability bounds. This still presents a substantial challenge, but also a possible hint at the nature of dynamical vacuum selection mechanisms.

We have computed the lifetime of the electroweak vacuum, $\tau$, as a function of the new physics parameters and of $m$ (in the $m^2 > 0$ case). In conjunction with the results in \cite{Khoury:2021zao}, these results will allow predictions for $\tau$ to be matched to a typical value of $m$. The proof-of-concept low-scale seesaw model we consider is, of course, only one example of the sort of TeV-scale new physics that can result in sharp metastability bounds.

The fact that the parameters in the Higgs potential all seem tuned to criticality may be a profound clue at a radically new kind of physics underlying the SM - perhaps these tunings, which resist individual explanation, all share a single origin. Thus, understanding the connections between these tunings might prove to be one of the first steps towards unveiling the nature of this new physics.
\begin{acknowledgments}
We thank Michael Geller for the discussions on the early stages of the project. We also thank Yannis Georis for discussions about low-scale neutrinos and their motivations.
T.S.'s contributions to this work were made possible by the Walter Benjamin Programme of the Deutsche Forschungsgemeinschaft (DFG, German Research Foundation) -- 512630918. Portions of this work were conducted in MIT's Center for Theoretical Physics and partially supported by the U.S. Department of Energy under Contract No.~DE-SC0012567. This project was also supported in part by the Black Hole Initiative at Harvard University, with support from the Gordon and Betty Moore Foundation and the John Templeton Foundation. The opinions expressed in this publication are those of the author(s) and do not necessarily reflect the views of these Foundations. S.T. was supported by the U.S. Department of Energy (DOE) Office of High Energy Physics under Grant Contract No. DE-SC0012567, and by the DOE QuantISED program through the theory consortium “Intersections of QIS and Theoretical Particle Physics” at Fermilab (FNAL 20-17). S.T. was additionally supported by the Swiss National Science Foundation - project n.~P500PT\_203156.
\end{acknowledgments}

\appendix
\section{Numerical Procedure for Lifetime Calculation}
\label{app:lifetime_calc}
As discussed in Sec.~\ref{sec:lifetime_constraints}, the bubble nucleation, or, equivalently, tunneling rate per unit volume is given by
\begin{equation}
    \frac{\Gamma}{V} = A e^{-S_E[\phi_I]}\,,
\end{equation}
where the instanton $\phi_I$ is a solution of the Euclidean equations of motion interpolating between the false vacuum and a point in field space of the same energy within the basin of another, lower energy vacuum. $S_E[\phi_I]$ is its Euclidean action, and $A$ is a dimensionful prefactor. $A$ contains the loop corrections to the decay rate and is of mass-dimension four. 

The coefficient $A$ arises from the integral over the fluctuations around the instanton and can be calculated in at least two ways. The first way is to integrate out fluctuations around the false vacuum configuration, and then compute the instanton using the resulting effective action. The second way is to find the instanton configuration around the original action, and to then perform the Gaussian integral over the fluctuations around the instanton background.

An important complication of the first approach is that the one-loop effective action itself is intractable in general; in most cases it is only known how to compute the effective potential, whereas the corrections to the kinetic term can only be obtained through a derivative expansion, c.f. Refs.~\cite{Moss:1985ve, Bodeker:1993kj, Salvio:2016mvj, Steingasser:2023gde}. Ultimately, when computing the effective action around the false vacuum, this derivative expansion only converges if its power counting parameter, the ratio of the effective mass of the scalar to that of the field being integrated out, is small. This implies that only those particles with a mass larger than the scalar can be integrated out in such a way, and, in particular, not the scalar itself. For a detailed discussion, see Ref. \cite{Andreassen:2016cvx}.

In our analysis, by explicitly checking the values of the effective masses of the fields at each point in our parameter space, we have verified that the derivative expansion for the effective action is well-behaved when integrating out all fields other than the Higgs itself. Therefore, we can perform the path integral over these fields and neglect the loop-generated derivative contributions to the effective action, relying on the effective potential Eq.~\eqref{eq:effpot} for our calculations. Using this potential, we then find the instanton $\phi_I$ and $S_E[\phi_I]$ using the shooting method.

To compute the dimensionful prefactor of the rate $\Gamma$, we compute the loop corrections due to the Higgs itself by performing the Gaussian integral over fluctuations around the instanton configuration. At one-loop, this is equivalent to evaluating a ratio of functional determinants,
\begin{equation}
\label{eq:detrat}
    A = \hspace{-0.2em} \left( \frac{\text{Det}(-\Delta_4 + W[\phi_I])}{\text{Det}(-\Delta_4 + W[\phi_\text{FV}])} \right)^{-\frac{1}{2}} \hspace{-0.4em} \equiv \left( \frac{\text{Det}( O_I)}{\text{Det}(O_{\rm FV})} \right)^{-\frac{1}{2}} \hspace{-1em}\,,
\end{equation}
where $\Delta_4$ is the four-dimensional Euclidean Laplacian, $W$ is the second derivative of the effective potential with respect to the Higgs field, and $\phi_\text{FV}$ is the false vacuum field configuration.

These determinants can be evaluated through the usual angular momentum decomposition. See Ref.~\cite{Steingasser:2022yqx} for details of this decomposition, as well as a pedagogical presentation of decay rate calculations in scalar field theory in general. The ratio of the functional determinants of these four-dimensional operators then decomposes into an infinite product of ratios of functional determinants of one-dimensional operators, each labeled by its corresponding angular momentum $l$:
\begin{align}
    \frac{\text{Det}(O_I)}{\text{Det}(O_\text{FV})} = & \prod_l \left(\frac{\text{Det}(O_I^l)}{\text{Det}(O_{\rm FV}^l)}\right)^{(l+1)^2} \\ 
    \equiv &\prod_l \left(\frac{\text{Det}(-D_{l,r} + W[\phi_I])}{\text{Det}(-D_{l,r} + W[\phi_\text{FV}])}\right)^{(l+1)^2}\,, \nonumber
\end{align}
where the arguments of the determinants on the left-hand side are the same as in Eqn.~\eqref{eq:detrat}, and
\begin{equation}
    D_{l,r} \equiv \partial_r^2 + \frac{3}{r} \partial_r - \frac{l(l+2)}{r^2}\,,
\end{equation}
is the $l$ mode of the decomposition of $\Delta_4$.

The operator corresponding to $l=0$ has a negative eigenvalue $\omega_-$, which, in principle, might seem to prevent the convergence of the corresponding Gaussian integral. The solution of this issue is well-understood~\cite{Coleman:1977py} and amounts to the replacement $\omega_- \to |\omega_-|$ up to a factor of $i$, which is absorbed by an imaginary part in the full expression for the decay rate. The operator corresponding to $l=1$ has four zero modes, corresponding to translations in Euclidean spacetime. These modes can be dealt with through the introduction of collective coordinates. The Jacobian of this transformation then gives rise to the prefactor of mass-dimension four. We follow the treatment of Ref.~\cite{Dunne:2005rt}, computing the zero-mode contribution in terms of the asymptotic behavior of the instanton. Moving forward, we denote by ${\rm Det}^\prime$ the functional determinant of $O_I$ with the zero modes removed in this manner.

The remaining factors can be evaluated numerically using the Gelfand-Yaglom method \cite{Gelfand:1959nq}, which allows one to compute such ratios of one-dimensional functional determinants of two operators in terms of the asymptotic behavior of their (not necessarily normalizable) zero modes. We denote the zero mode of the numerator operator $\psi_\text{instanton}^l$ and that of the denominator operator $\psi_\text{free}^l$. Having found any two such functions, the ratio of the functional determinants can be shown to be given by
\begin{equation}
\label{eq:GY}
 \frac{\text{Det}^\prime (O_I^l)}{\text{Det}(O_{\rm FV}^l)} = \lim_{r\to\infty}\frac{\psi_\text{free}^l(0)}{\psi_\text{instanton}^l(0)} \cdot \frac{\psi_\text{instanton}^l(r)}{\psi_\text{free}^l(r)}\,.
\end{equation}
Due to the linearity of the zero-mode equations, we are free to rescale these modes s.t. they coincide in the limit $r\to 0$, thus eliminating the first factor in Eq.~\eqref{eq:GY}. To improve numerical stability during the analysis of their asymptotics, we furthermore define the ratio $\rho_l$ as
\begin{equation}
    \rho_l(r) \equiv \frac{\psi_\text{instanton}^l(r)}{\psi_\text{free}^l(r)}\,.
\end{equation}
Specifying the Higgs potential with $m^2 > 0$, $\phi_\text{FV} = 0$, $\psi^l_\text{free}$ satisfies the equation
\begin{equation}
    (-\partial^2_r - \frac{3}{r}\partial_r + \frac{l(l+2)}{r^2} + W[0])\psi^l_\text{free}(r) = 0\,,
\end{equation}
which admits the analytic solution
\begin{equation}
    \psi^l_\text{free}(r) = \frac{2^{l+1}(l+1)!}{r}I_{l+1}(r W[0])\,,
\end{equation}
where $I_{l+1}$ is the modified Bessel function. This allows us to construct a differential equation for $\rho_l(r)$ using the zero-mode equation for $\psi^l_\text{instanton}$,
\begin{align}
    0=&-\rho^{ \prime \prime}_l(r) - 2 \rho^{\prime}_l(r) \frac{\psi^{l \prime}_\text{free}(r)} {\psi^l_\text{free}(r)} \\ &- \frac{3}{r}\rho_l^{\prime}(r) + \rho_l(r) \left( W[\phi_I] - W[0]\right)\,.\nonumber
\end{align}
This equation can be solved numerically, and we may express the desired ratio of functional determinants as
\begin{equation}
    \frac{\text{Det}^\prime(O_I)}{\text{Det}(O_{\rm FV})} = \prod_l \left(\lim_{r \to \infty}\rho^l(r)\right)^{(l+1)^2}\,.
\end{equation}
This infinite product is divergent and must be regularized and renormalized. As detailed in Ref. \cite{Dunne:2005rt}, the large $l$ asymptotic behavior of the product is well-described analytically by the WKB approximation. The renormalized product can then be calculated by matching the product onto this analytic form at large $l$, performing the regularization and renormalization at the level of the analytic expression and finally truncating and matching the product at a maximum value of $l$ denoted by $l_\text{max}$. In our numerical calculation we take $l_\text{max} = 50$, which we have confirmed numerically to be large enough for the resulting error to be negligible.

Our final result for the contribution of the one-loop corrections can be presented most conveniently in terms of new variables $B$ and $S_E^{(1)}$ defined through
\begin{equation}
    B e^{-S_E^{(1)}} \equiv A = \left( \frac{\text{Det}(O_I)}{\text{Det}(O_{\text{FV}})} \right)^{-\frac{1}{2}}\,,
\end{equation}
where 
\begin{align}
     B =& \left(2 \pi \phi_\text{inf} (\partial^2_r \phi_I(0)) \right)^2, \ \ {\rm with} \\ 
     \phi_\text{inf} \equiv & \lim_{r \to \infty} \frac{r \phi_I(r)}{\sqrt{W[0]} K_1\left(r \sqrt{W[0]} \right)}\,.
\end{align}
Here, $K_1$ denotes the modified Bessel function of the second kind. The second derivative of the instanton at the origin can be evaluated using the equations of motion and initial conditions for the instanton. For $S_E^{(1)}$, we find
\begin{widetext}
\begin{align}
\label{eq:rate}
    S_E^{(1)} &= \frac{1}{2} \log(|R_0|)  + \frac{1}{2} \sum_{l=2}^{l_\text{max}} \log(R_l) - \frac{(l_\text{max}+1)(l_\text{max}+2)}{8} \int_0^\infty \text{d}r \left[r W_b\right]\\
    &+ \frac{\log l_\text{max}}{16} \int_0^\infty \text{d}r \left[ r^3 W_b (W_b + 2 W[0]) \right] - \frac{1}{16} \int_0^\infty \text{d} r \left[ \left(1 + \frac{\mu r}{2}\right) r^3 W_b (W_b + 2 W[0]) \right]\,,\nonumber
\end{align}
\end{widetext}
where $R_l$ is defined as $(\lim_{r \to \infty} \rho_l(r))^{(l+1)^2}$, $W_b$ is $W[\phi_I] - W[0]$, and $\mu$ is the renormalization scale. To ensure the convergence of our perturbative calculation, we choose $\mu=\mu_I\sim \phi_I (0)$.

\begin{figure}
    \centering
    \includegraphics[width=\columnwidth]{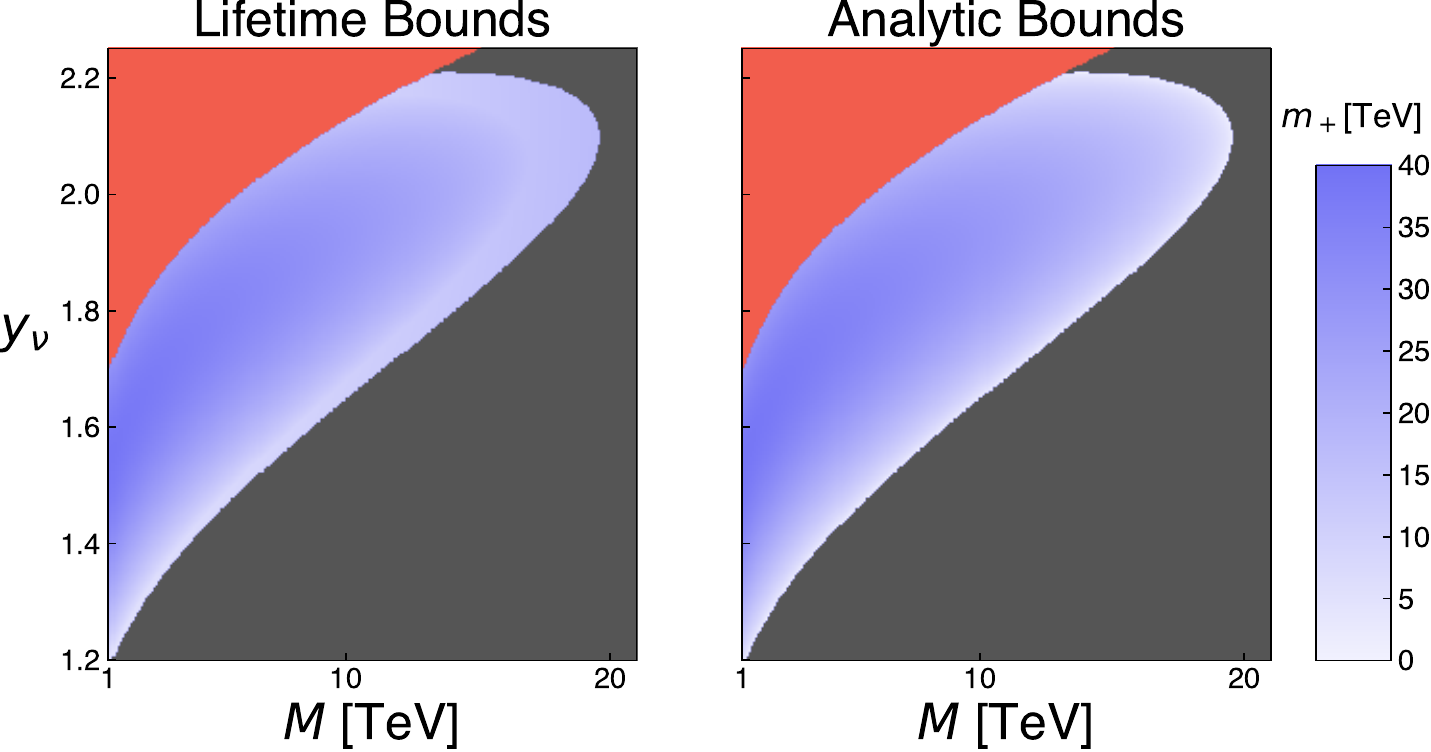}
    \caption{A comparison of the upper metastability bound on $m^2$, $m_+^2$, as derived from the lifetime computation (left) and as derived from the analytic computation (right). As before, the region shaded blue is the region where a bound is present, and lighter blues correspond to stronger bounds (see the colorbar for reference). Red denotes couplings becoming non-perturbative, and grey means that metastability is violated for any value of $m^2$.}
    \label{fig:fig8}
\end{figure}

As a crosscheck of this calculation, we compare the analytic value of $m_+^2$ to the value of $m^2$ for which the lifetime becomes infinite. In particular, since the existence of a metastable vacuum manifests in our calculation as the existence of an instanton configuration, $m_+^2$ should manifest in the lifetime computation as the smallest value of $m^2$ for which there does not exist an instanton. There is a small conceptual difference between the two bounds; the analytic bound detects when the potential's second minimum ceases to exist, whereas the instanton ceases to exist when there are still two minima, but they become degenerate. In practice, we find that this difference is insignificant.
We show $m_+^2$, as derived from the lifetime computation and from the analytic computation, in Fig. \ref{fig:fig8}.
The left panel shows $m_+$ as derived from the lifetime computation, and the right panel (identical to Fig. \ref{fig:fig3} but included here for convenience) shows $m_+$ as derived from the analytic bound.
We can see that we observe good agreement between the two calculations.
\bibliography{ref}
\end{document}